\documentclass[pra,aps,10pt,twocolumn,showpacs,nofootinbib]{revtex4-1}

\usepackage{graphicx,amsmath,amsfonts,bm}

\newcommand{\abs}[1]{\left| #1 \right|} 
\newcommand{\ii}{\mathrm{i}}	        
\newcommand{\ee}{\mathrm{e}}    	
\newcommand{\dd}{\mathrm{d}}		
\newcommand{\Braket}[3]{\left< #1 \vphantom{#2#3} \right| #2 \vphantom{#1#3}
  \left| #3 \vphantom{#1#2} \right>}    
\renewcommand{\vec}[1]{\boldsymbol{#1}}	

\begin{document}

\title{Macroscopic quantum tunneling of Bose-Einstein condensates with
long-range interaction}

\author{Kai Marquardt}
\author{Pascal Wieland}
\author{Rolf H\"afner}
\author{Holger Cartarius}
\author{J\"org Main}
\author{G\"unter Wunner}
\affiliation{Institut f\"ur Theoretische Physik 1, Universit\"at Stuttgart,
  70550 Stuttgart, Germany}
\date{\today}

\begin{abstract}
The ground state of Bose-Einstein condensates with attractive particle
interaction is metastable.  One of the decay mechanisms of the
condensate is a collapse by macroscopic quantum tunneling, which can
be described by the bounce trajectory as solution of the
time-dependent Gross-Pitaevskii equation in imaginary time.  For
condensates with an electromagnetically induced gravity-like
interaction the bounce trajectory is computed with an extended
variational approach using coupled Gaussian functions and simulated
numerically exact within the mean-field approach on a space-time
lattice.  It is shown that the variational computations converge very
rapidly to the numerically exact result with increasing number of
Gaussians.  The tunneling rate of the condensate is obtained from the
classical action and additional parameters of the bounce trajectory.
The converged variational and numerically exact results drastically
improve by several orders of magnitude the decay rates obtained
previously with a simple variational approach using a single
Gaussian-type orbital for the condensate wave function.
\end{abstract}


\maketitle

\section{Introduction}
\label{sec:intro}
The macroscopic occupation of the bosonic ground state of ultracold
quantum gases has been predicted by Bose and Einstein.
At least since the first realization of Bose-Einstein condensates
(BECs) in 1995 \cite{And95a,Bra95a,Dav95a} they are a central part of
experimental and theoretical atomic physics.
In a cold diluted Bose gas the particles interact via short-range
contact interactions determined by the s-wave scattering length
$a$, which can be experimentally varied with the help of Feshbach
resonances.
In addition long-range interactions exist, e.g., in dipolar
condensates \cite{Lahaye09}, or in condensates with an attractive $1/r$
interaction, which have been proposed to be electromagnetically
induced with a setup of six laser triads \cite{ODe00}.
The condensates are typically held in a harmonic trap, however, in
case of an attractive $1/r$ interaction the BEC can be self-trapped
\cite{ODe00,Pap07}, which means that no external trap is needed.

When the scattering length of the contact interaction is varied, the
stable ground state plus an unstable excited state of a BEC with
attractive $1/r$ interaction are created in a tangent bifurcation at a
critical scattering length $a=a_{\rm cr}$ \cite{Pap07,Car08a}.
Below the critical scattering length the contact interaction is so
attractive that the condensate collapses and no stationary state does
exist \cite{Car08b}.
The collapse of the condensate, e.g., induced by thermal excitations
\cite{Hue99,Hue03,Jun12b,Jun12c}, is also possible at scattering
lengths above the bifurcation point.
However, even without thermal excitations, i.e., at zero temperature
$T=0$, the ground state of the condensate is only a metastable state
and can collapse by macroscopic quantum tunneling as long as the
contact interaction is attractive.
The process can be described by using the bounce trajectory as
outlined by Stoof \cite{Sto97} and by Freire and Arovas \cite{Fre99}.

The aim of the present paper is a thorough investigation and
computation of the bounce trajectory of a BEC.
It has been shown that the stationary states of the GPE can be
excellently described with an extended variational approach using
coupled Gaussian functions \cite{Rau10a,Rau10b}.
Here we will demonstrate that computations with coupled Gaussians are
also a full-fledged alternative to simulations on a space-time lattice
to obtain the bounce trajectory which describes the macroscopic
quantum tunneling of the BEC.
We apply the formalism to a self-trapped BEC with an attractive $1/r$
interaction.
On the one hand the symplicity of this spherically symmetric system
allows for an intuitive access to the bounce trajectory similar to the
one-dimensional potential picture used by Stoof \cite{Sto97} and
Freire et al.\ \cite{Fre99,Fre97}.
On the other hand we can show that it is possible to include
long-range interactions into the instanton formalism for BECs.
The theory presented in this article can be applied to more realistic
dipolar condensates \cite{Lahaye09} as well since their description does
not contain a conceptional difference.

The paper is organized as follows.
The basic concepts are briefly outlined in Sec.~\ref{sec:basics}.
In Sec.~\ref{sec:gauss_calc} the time-dependent variational principle
(TDVP) is applied to an ansatz of coupled Gaussians for the wave
functions in imaginary time.
Periodic solutions of the equations of motion for the variational
parameters are obtained by application of a multi-shooting algorithm,
which converge with increasing period to the bounce trajectory.
In Sec.~\ref{sec:grid_calc} the method for exact numerical simulations
on space-time lattices is elaborated.
The results of both methods are presented and compared in
Sec.~\ref{sec:res}.
Concluding remarks and an outlook are given in Sec.~\ref{sec:conc}.

\section{Basic concepts}
\label{sec:basics}
For condensates without long-range interactions the macroscopic
quantum tunneling has been investigated by Stoof \cite{Sto97} using a
variational approach with a single Gaussian function as a simple
approximation.
For a self-trapped BEC with an attractive $1/r$ interaction instead of
an external harmonic potential the ideas outlined in Ref.~\cite{Sto97}
can briefly be summarized as follows.
Starting from the many-body Schr\"odinger equation a mean-field
approach leads to the time-dependent nonlinear Gross-Pitaevskii
equation (GPE)
\begin{equation}
 \ii\frac{\dd}{\dd t} \psi(\vec{r},t) = \left[ - \Delta +
   V_\mathrm{c}(\vec{r},t) + V_\mathrm{u}(\vec{r},t)\right] \psi(\vec{r},t) 
\label{eq:GPE}
\end{equation}
with the contact and long-range potentials
\begin{align}
\label{eq:Vc}
 V_\mathrm{c}(\vec{r},t) &= 8\pi N a \abs{\psi(\vec{r},t)}^2 \; ,\\
\label{eq:Vu}
 V_\mathrm{u}(\vec{r},t) &= -2N \int \dd^3 r'
   \frac{\abs{\psi(\vec{r'},t)}^2}{\abs{\vec r  - \vec r'}} \; ,
\end{align}
where $a$ is the scattering length, $N$ is the number of bosons, and
``natural'' units \cite{Pap07} were used.
Lengths are measured in units of $a_u=\hbar^2/(mu)$, energies in units
of $E_u=\hbar^2/(2ma_u^2)$, and times in units of $t_u=\hbar/E_u$,
where $u$ determines the strength of the atom-atom interaction
\cite{ODe00}, and $m$ is the mass of one boson.
Exploiting the scaling properties presented in Ref.~\cite{Pap07} and
introducing the scaled variables
\begin{equation}
 (\tilde{\vec{r}}, \tilde a, \tilde t, \tilde\psi) = 
 (N\vec{r}, N^2a, N^2\tilde t, N^{-3/2}\psi)
\label{eq:Nscal}
\end{equation}
the GPE keeps the form given in Eq.~\eqref{eq:GPE}, however, with the
particle number in the potentials $V_{\rm c}$ and $V_{\rm u}$ in
Eqs.~\eqref{eq:Vc} and \eqref{eq:Vu} formally set to $N=1$.
Note that the scaled GPE for the self-trapped BEC only depends on one
parameter, viz.\ the scaled scattering length.
In the following we use the scaled variables unless otherwise stated
and omit the tilde.

\begin{figure}
\includegraphics[width=0.95\columnwidth]{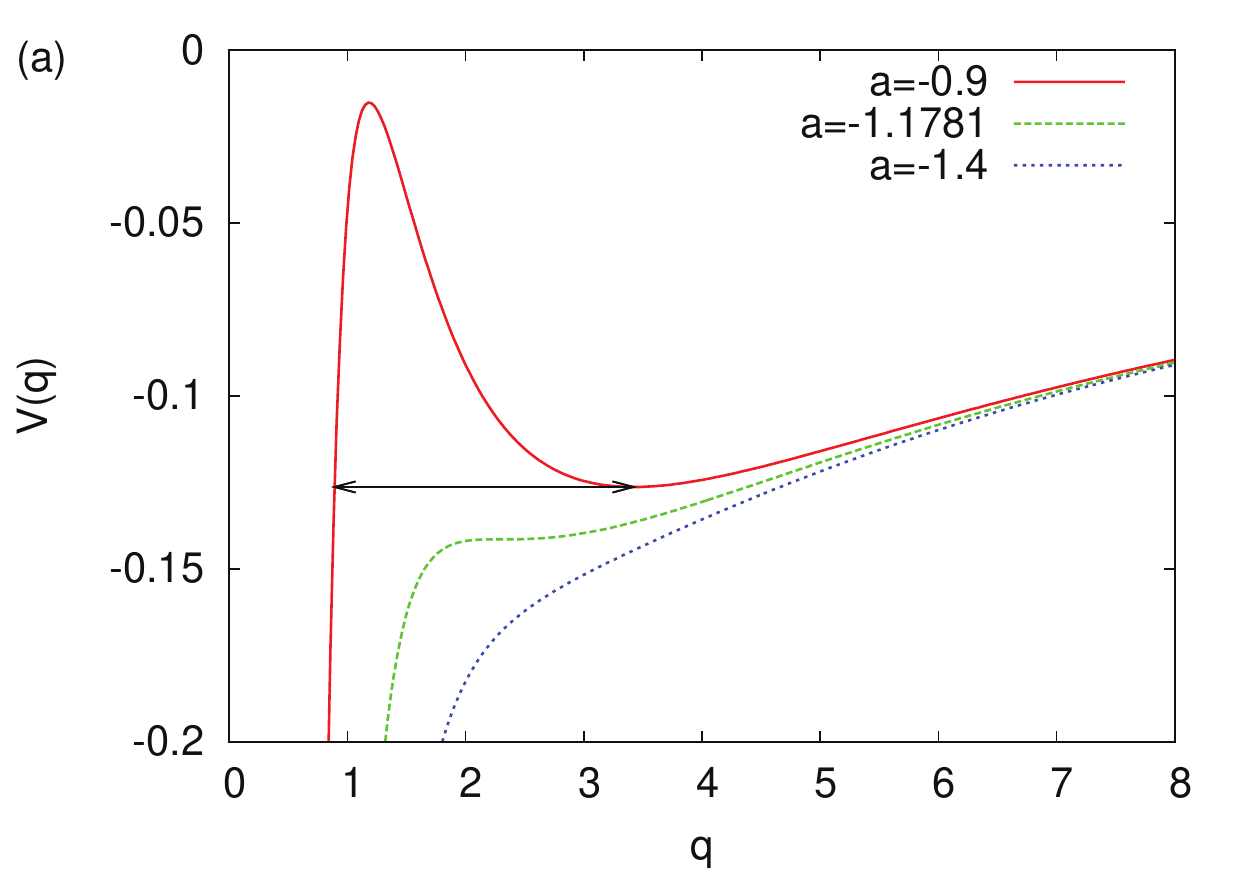}
\includegraphics[width=0.65\columnwidth]{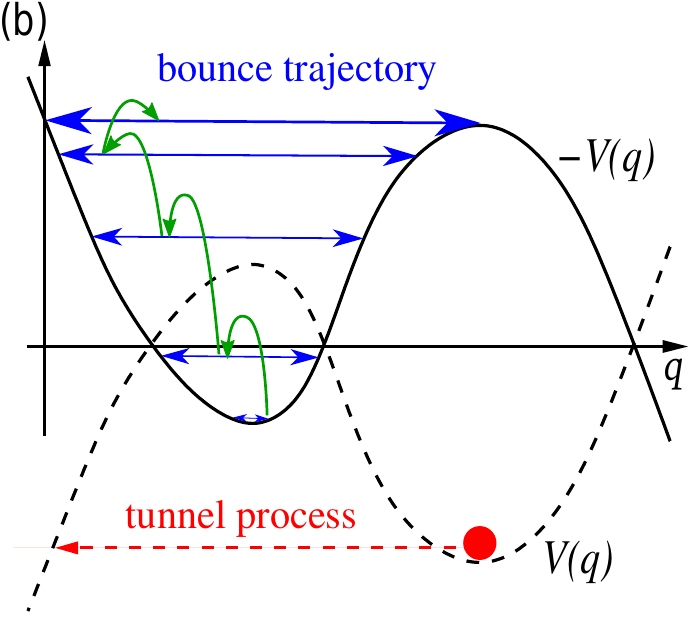}
\caption{(Color online)
 (a) Potential $V(q)$ of a self-trapped BEC with attractive $1/r$
 interaction at various values of the scattering length $a$.
 (b) Scheme of the inversion of the potential and the periodic orbits
 approaching the bounce trajectory.}
\label{fig1}
\end{figure}

Using a Gaussian-type orbital
\begin{equation}
 \psi(r,t) = 4\left[\frac{2}{3\pi q(t)^2}\right]^{3/4}
   \exp\left\{\left[-\frac{4}{3q(t)^2}
     +\ii\frac{p(t)}{2q(t)}\right]r^2\right\}
\label{eq:GTO}
\end{equation}
normalized to $||\psi||^2=1$ as a simple variational ansatz for the
condensate wave function the TDVP can be applied, and it can be shown
\cite{Car08b} that the parameters $q$ and $p$ obey the canonical
equations of motion
\begin{equation}
 \dot q = \frac{\partial H}{\partial p} \; , \quad
 \dot p = -\frac{\partial H}{\partial q} \; ,
\end{equation}
with the Hamiltonian $H(q,p)=p^2+V(q)$ and the potential
\begin{equation}
 V(q) = \frac{9}{4q^2} + \sqrt{\frac{3}{\pi}}\frac{3a}{2q^3}
   - \sqrt{\frac{3}{\pi}}\frac{1}{q} \; .
\label{eq:Vq}
\end{equation}
Important features of the BEC can be simply obtained, at least
qualitatively, from the potential $V(q)$ shown in Fig.~\ref{fig1}(a).
For scattering lengths $a<a_{\rm cr}=-3\pi/8=-1.1781$ the potential
$V(q)$ does not possess any stationary points, i.e., the condensate
collapses.
For $a_{\rm cr}<a<0$ two stationary points exist, the local minimum at
$q=q_{\min}$ and maximum at $q=q_{\max}$ of the potential indicate the
stable ground state and the unstable excited state, with $V(q_{\min})$
and $V(q_{\max})$ the mean-field energies
$E_{\rm mf}=\langle\psi|-\Delta+\frac{1}{2}(V_{\rm c}+V_{\rm u})|\psi\rangle$
of the two states, respectively.

From Fig.~\ref{fig1}(a) it is obvious that the potential $V(q)$ at
small values of $q$ drops below the mean-field energy of the ground state.
The path through the potential barrier connecting the position of the
ground state and the collapsing region [see the arrow in
Fig.~\ref{fig1}(a)] is classically forbidden.
In quantum mechanics, however, the barrier can be crossed by a
tunneling process, which is described by propagation along that path,
called the bounce trajectory, in imaginary time.
The decay rate of the metastable ground state depends on parameters of
the bounce trajectory.

In the mean-field approach the many-body ground state is the product
of $N$ identical one-particle wave functions, and thus the collapse of
a BEC requires the collective tunneling of all $N$ bosons.
As a consequence for macroscopic quantum tunneling an effective Planck
constant $\hbar/N$ must be used in the formula for the decay rate
$\Gamma_0$.
The final result derived by Stoof \cite{Sto97} reads
\begin{align}
 \Gamma_0 = \sqrt{\frac{N m \omega_0 v_0^2}{\pi \hbar}}
 \ee^{-\frac{N}{\hbar} S_{\rm b}}
\label{eq:Gamma0}
\end{align}
where
\begin{equation}
 S_{\rm b} = 2\int_{q_{\rm b}}^{q_{\min}}\sqrt{V(q)-V(q_{\min})}\dd q
\end{equation}
is the Euclidean action of the bounce trajectory (with the turning
points $q_{\rm b}$ and $q_{\min}$) in imaginary time $\ii t\to\tau$
running in the inverted potential \eqref{eq:Vq}, i.e., $V(q)\to -V(q)$
as illustrated in Fig.~\ref{fig1}(b).
The parameter $\omega_0$ is the frequency of oscillations in the
potential minimum around $q\approx q_{\min}$ given by
\begin{equation}
 \omega_0 = \frac{1}{m} V''(q_{\min}) \; ,
\label{eq:omega0}
\end{equation}
and $v_0$ is defined by the condition that the bounce trajectory
approaches the maximum of the inverted potential $-V(q)$ [see
Fig.~\ref{fig1}(b)] in the limit $\tau\to\pm\infty$ as
\begin{equation}
 q(\tau) \sim q_{\min} - \frac{v_0}{\omega_0}\ee^{-\omega_0|\tau|} \; .
\label{eq:v0_q}
\end{equation}

The computation of the decay rate $\Gamma_0$ in Eq.~\eqref{eq:Gamma0}
with the parameters $S_{\rm b}$, $\omega_0$, and $v_0$ of the bounce
trajectory as explained above is only a rough approximation for the
following two reasons.
Firstly, the ansatz with a single Gaussian function in
Eq.~\eqref{eq:GTO} cannot describe the true condensate wave function.
The variational solutions significantly deviate from those found in
full numerical simulations even for the stationary ground and excited
state \cite{Pap07,Car08a}.
Secondly, the true solution of the GPE \eqref{eq:GPE} can fluctuate
around both the stationary ground state and the time-dependent bounce
trajectory.
The modes of these fluctuations are described, in leading order, by
the Bogoliubov-de Gennes equations \cite{Pit03}.
In Eq.~\eqref{eq:Gamma0} only fluctuations along the direction of the
path are considered, i.e., for a system with only one degree of
freedom.
The true condensate wave function has many degrees of freedom, which
allows for fluctuations perpendicular to the direction of the path.
The fluctuation determinant resulting from those additional degrees of
freedom is not accounted for by Eq.~\eqref{eq:Gamma0}.

The method proposed by Stoof \cite{Sto97} requires the description of
the bounce trajectory with canonical variables and the Hamiltonian
formalism.
Such a description is, however, restricted to a variational ansatz for
the condensate wave function with a single Gaussian function as in
Eq.~\eqref{eq:GTO}, and thus any improvement of the simple variational
approach is a nontrivial task.

A decisive step for the improvement of the bounce trajectory approach
was achieved by Freire and Arovas \cite{Fre99}.
They found a way to compute the bounce process beyond the
interpretation in terms of a classical trajectory by applying the
instanton formalism to a BEC in a harmonic trap without long-range
interaction.
The bounce trajectory was simulated numerically on a space-time
lattice without any restriction to the wave function.

To solve the GPE \eqref{eq:GPE} in imaginary time $\ii t\to\tau$ the
two fields $\psi(\vec r,\tau)$ and $\bar\psi(\vec r,\tau)$ are
introduced, which obey the equations
\begin{equation}
 -\frac{\dd}{\dd\tau}\psi(\vec r,\tau) = H\psi(\vec r,\tau) \; , \;
  \frac{\dd}{\dd\tau}\bar\psi(\vec r,\tau) = H\bar\psi(\vec r,\tau) \; ,
\end{equation}
with
\begin{align}
\label{eq:H}
 H(\vec r,\tau) &= -\Delta + V_{\rm c}(\vec r,\tau)
  + V_{\rm u}(\vec r ,\tau)-\mu \; ,\\
\label{eq:Vc_tau}
 V_{\rm c}(\vec r,\tau) &= 8\pi N a
 \bar\psi(\vec r ,\tau)\psi(\vec r ,\tau) \; ,\\
\label{eq:Vu_tau}
 V_{\rm u}(\vec r ,\tau) &= -2N \int \dd^3 r'
   \frac{\bar\psi(\vec r',\tau)\psi(\vec r',\tau)}{|\vec r - \vec r'|} \; .
\end{align}
Note that $\mu$ is a free parameter in Eq.~\eqref{eq:H} and the
function $\bar\psi(\vec r,\tau)$ in imaginary time plays the role of
the complex conjugate $\psi^\ast(\vec r,t)$ in real time dynamics.
The probability density is given by $\bar\psi\psi$ and the norm
defined by $\int\bar\psi(\vec r ,\tau)\psi(\vec r ,\tau)\dd^3 r$
is conserved.
The bounce trajectory (with real valued $\psi$ and $\bar\psi$) must
fulfill the boundary conditions \cite{Fre99,Fre97}
\begin{align}
\label{eq:psi_sym}
\bar\psi(\vec r,\tau)&=\psi(\vec r,-\tau) \; ,\\
\label{eq:psi_lim}
\lim_{\tau\to\pm\infty}\bar\psi(\vec r,\tau)&=
\lim_{\tau\to\pm\infty}\psi(\vec r,\tau)\equiv\psi_{\rm g}(\vec r) \; ,
\end{align}
with $\psi_{\rm g}(\vec r)$ the wave function of the stationary ground
state.
The Euclidean action of a periodic trajectory in imaginary time with
period $\beta$ is given by
%
\begin{equation}
 S_{\ee} = \int_{-\beta/2}^{\beta/2} \dd\tau \int\dd^3r
  \left\{ \bar\psi \frac{\dd}{\dd\tau} \psi - \mathcal{H} \right\} \; ,
\label{eucAction}
\end{equation}
with the associated Hamiltonian density
\begin{equation}
 \mathcal{H} = -\nabla \bar\psi \nabla \psi + 4\pi a
  \left(\bar\psi\psi \right)^2 - \int \dd^3 r'
  \frac{\psi' \bar\psi' \psi\bar\psi}{|\vec r - \vec r'|} \; .
\label{hamDens}
\end{equation}
Due to the symmetry \eqref{eq:psi_sym} of $\psi$ and $\bar\psi$ and
the behavior of $S_{\ee}[q_{\mathrm b}]$ and $S_{\ee}[q_0]$ the difference
$\Delta S_{\ee}=S_{\ee}[q_{\mathrm b}]-S_{\ee}[q_0]$
between the actions of the bounce trajectory and the fixed point is
obtained as \cite{Fre99}
\begin{equation}
 \Delta S_{\ee} = \int_0^{\beta/2} \dd \tau \int \dd^3 r
 \left\{ \bar\psi \frac{\dd}{\dd\tau} \psi - \psi \frac{\dd}{\dd\tau}
 \bar\psi \right\} \; .
\label{eq:Se}
\end{equation}
Within the mean-field approach the formalism allows, in principle, for
the exact computation of the bounce trajectory and related parameters
needed as input for the calculation of the decay rate of the BEC in
Eq.~\eqref{eq:Gamma0} without resort to the ansatz \eqref{eq:GTO} and
any canonical variables.

\section{Calculation of the bounce trajectory}
\label{sec:calc}
The computation of the exact bounce trajectory which describes within
the mean-field approach the macroscopic quantum tunneling of a
metastable BEC is the central issue of this paper.
We first introduce in Sec.~\ref{sec:gauss_calc} the extended
variational approach with coupled Gaussian functions, and then in
Sec.~\ref{sec:grid_calc} the numerical simulations on space-time
lattices.
The derivations are presented for a self-trapped BEC with attractive
$1/r$ interaction, where the condensate wave function is spherically
symmetric.

\subsection{Variational approach with coupled Gaussians}
\label{sec:gauss_calc}
The variational ansatz with a single Gaussian function in
Eq.~\eqref{eq:GTO} cannot exactly reproduce the true condensate wave
function.
However, it has been shown that an extended variational approach with
coupled Gaussians yields excellent results for both the stationary
ground and excited state of the BEC \cite{Rau10b}.
Only very few (about three to five) coupled Gaussians are sufficient
to achieve convergence of the wave function, the mean-field energy,
and the chemical potential to those obtained in numerical computations
on grids.
We therefore will use a variational approach with coupled Gaussians
also for the calculation of the bounce trajectory.
For the dynamical calculations in imaginary time we write both
functions $\psi$ and $\bar\psi$ as superpositions of $N_g$ Gaussian
functions,
\begin{subequations}
\begin{align}
 \psi(r,\tau) &= \sum_{k=1}^{N_g}
  \ee^{-\left[ A_k(\tau) r^2 + \gamma_k(\tau) \right]}
  \equiv \sum_{k=1}^{N_g} g_k \; ,\\
 \bar\psi(r,\tau) &= \sum_{k=1}^{N_g}
  \ee^{-\left[ \bar A_k(\tau) r^2 + \bar\gamma_k(\tau) \right]}
  \equiv \sum_{k=1}^{N_g} \bar g_k \; .
\end{align}
\label{eq:coupled_gauss}
\end{subequations}
The $4N_g$ time-dependent variational parameters can be written as
two vectors
\begin{subequations}
\begin{align}
 \vec z &= (A_1,\dots,A_{N_g},\gamma_1,\dots,\gamma_{N_g}) \; ,\\
 \bar{\vec z} &= (\bar A_1,\dots,\bar A_{N_g},
   \bar\gamma_1,\dots,\bar\gamma_{N_g}) \; .
\end{align}
\end{subequations}
The equations of motion for the variational parameters $\vec z$ and
$\bar{\vec z}$ can be obtained using the TDVP of Dirac, Frenkel, and
McLachlan \cite{Dir30,McL64}, which must be modified and adopted in a
special way for the dynamics in imaginary time.
In imaginary time the functional
\begin{equation}
 I = \int\dd^3r \left[\bar\phi - H \bar\psi] [\phi + H \psi\right]
\label{eq:I}
\end{equation}
must be minimized with respect to $\phi=\dot\psi$ and
$\bar\phi=\dot{\bar\psi}$, with the Hamiltonian $H$ given in
Eq.~\eqref{eq:H}.
The dots now mark derivatives $\dd/\dd\tau$.
With
\begin{equation}
 \delta\phi = \sum_{k=1}^{2N_g}
  \left\{ \frac{\partial\psi}{\partial z_k}
  \delta\dot z_k \right\} \; , \quad
 \delta\bar\phi = \sum_{k=1}^{2N_g}
  \left\{\frac{\partial\bar\psi}{\partial\bar z_k}
  \delta\dot{\bar z}_k \right\} \; ,
\end{equation}
we obtain the condition
\begin{align}
 \delta I &= \int\dd^3r \left\{\delta\bar\phi\left[\phi + H\psi\right]
  + \delta\phi\left[\bar\phi - H\bar\psi\right]\right\} \nonumber \\
 &= \int\dd^3r \sum_{k=1}^{2N_g} \left\{ \left[\left(\phi + H\psi\right)
   \frac{\partial \bar{\psi}}{\partial \bar{z}_k} \right]
   \delta\dot{\bar z}_k \right. \nonumber \\
  & \left. \qquad + \left[ \left( \bar\phi - H\bar{\psi}\right)
   \frac{\partial \psi}{\partial z_k}\right]\delta\dot z_k \right\} = 0 \; ,
\label{eq:delta_I}
\end{align}
and thus, because all $\delta\dot z_k$ and $\delta\dot{\bar z}_k$ are
independent
\begin{subequations}
\label{eq:TDVP_result}
\begin{align}
 \int\dd^3r \left[\left( \dot\psi + H \psi\right)
 \frac{\partial \bar{\psi}}{\partial\bar{z}_k} \right] &= 0 \; ,\\
 \int\dd^3r \left[ \left( \dot{\bar\psi} - H\bar{\psi}\right)
 \frac{\partial \psi}{\partial z_k}\right] &= 0 \; ,
\end{align}
\end{subequations}
for $k=1,\dots,2N_g$.
Using the ansatz \eqref{eq:coupled_gauss} with coupled Gaussians and
the Hamiltonian \eqref{eq:H} in Eq.~\eqref{eq:TDVP_result} a
calculation similar to that in Ref.~\cite{Rau10a} yields the equations
of motion for the variational parameters
\begin{subequations}
\label{eq:eom}
\begin{align}
 \dot A_k &= -4A_k^2 + v_k^{(2)} \; , \\
 \dot\gamma_k &= 6A_k + v_k^{(0)} - \mu \; , \\
 \dot{\bar A}_k &= 4\bar{A}_k^2 - \bar{v}_k^{(2)} \; , \\
 \dot{\bar\gamma}_k &= -6\bar{A}_k - \bar{v}_k^{(0)} + \mu \; ,
\end{align}
\end{subequations}
with $k=1,\dots,N_g$.
As in Eq.~\eqref{eq:H} $\mu$ is a free parameter, which becomes the
chemical potential for stationary states.
The parameters $v_k^{(2)}$, $v_k^{(0)}$, $\bar v_k^{(2)}$ and $\bar v_k^{(0)}$
in Eq.~\eqref{eq:eom} are obtained by solving the two linear systems
of equations
\begin{subequations}
\begin{align} 
 \sum_{k=1}^{N_g} v_k^{(0)} [1]_{lk} + \sum_{k=1}^{N_g} v_k^{(2)} [r^2]_{lk}
  &= \sum_{k=1}^{N_g} [V]_{lk} \; ,\displaybreak[0]\\
 \sum_{k=1}^{N_g} v_k^{(0)} [r^2]_{lk} + \sum_{k=1}^{N_g} v_k^{(2)} [r^4]_{lk}
  &= \sum_{k=1}^{N_g} [r^2V]_{lk} \; ,
\end{align}
\end{subequations}
and
\begin{subequations}
\begin{align} 
 \sum_{k=1}^{N_g} \bar v_k^{(0)} [1]_{kl} + \sum_{k=1}^{N_g} \bar v_k^{(2)}
  [r^2]_{kl} &= \sum_{k=1}^{N_g} [V]_{kl} \; ,\\
 \sum_{k=1}^{N_g} \bar v_k^{(0)} [r^2]_{kl} + \sum_{k=1}^{N_g} \bar v_k^{(2)}
  [r^4]_{kl} &= \sum_{k=1}^{N_g} [r^2V]_{kl} \; ,
\end{align}
\end{subequations}
with $V=V_{\rm c}+V_{\rm u}$ and the notation
\begin{equation}
 [O]_{lk} = \langle\bar g_l|O|g_k\rangle = \int\dd^3r \bar g_l O g_k \; . 
\end{equation}
All required integrals can be calculated analytically and are given in
Appendix~\ref{sec:app_A}.

The equations of motion \eqref{eq:eom} are a set of $4N_g$ ordinary
differential equations, which can be integrated, e.g., with a
Runge-Kutta method.
Because the period $\beta$ of the bounce trajectory is infinite we
start searching for periodic orbits, with the boundary conditions
$\bar\psi(\tau=0)=\psi(\tau=0)$ and
$\bar\psi(\tau=\beta/2)=\psi(\tau=\beta/2)$, near the excited
state, as illustrated in Fig.~\ref{fig1}(b), and then increase
$\beta$.
In the limit $\beta\to\infty$ the periodic orbits converge towards
the bounce trajectory.
The periodic orbit search could be performed, in principle, by varying
the initial conditions of the variational parameters at $\tau=0$ with
$\bar\psi(0)=\psi(0)$ in such a way, that
$\bar\psi(\beta/2)=\psi(\beta/2)$ is fulfilled at the other turning
point.
Numerically, however, this is an impossible task because the equations
of motion are extremely unstable.
The reason is that, e.g., for the ground state all stable modes of the
Bogoliubov spectrum in real time become unstable modes in imaginary
time, and thus all perturbations of the ground state (or a trajectory)
increase exponentially in imaginary time.
To overcome this problem we use a multi-shooting algorithm, where the
trajectory is divided into small segments.
Using $M$ segments the total set of $4MN_g+1$ parameters [including the
parameter $\mu$ in the equations of motion \eqref{eq:eom}] can be
determined to fulfill the $4MN_g$ continuity conditions of the
variational parameters and the condition $\int\dd^3r\bar\psi\psi=1$
for normalization by using a multidimensional Newton's method.
Because the norm of $\bar\psi\psi$ is conserved, the total set of
parameters can be slightly reduced by introducing the parameters
$\tilde\gamma_k=\gamma_k-\gamma_1$ and
$\tilde{\bar\gamma}_k=\bar\gamma_k-\bar\gamma_1$ for $k=2,\dots,N_g$
instead of $\gamma_k$ and $\bar\gamma_k$ for $k=1,\dots,N_g$.

\begin{figure}
\includegraphics[width=0.9\columnwidth]{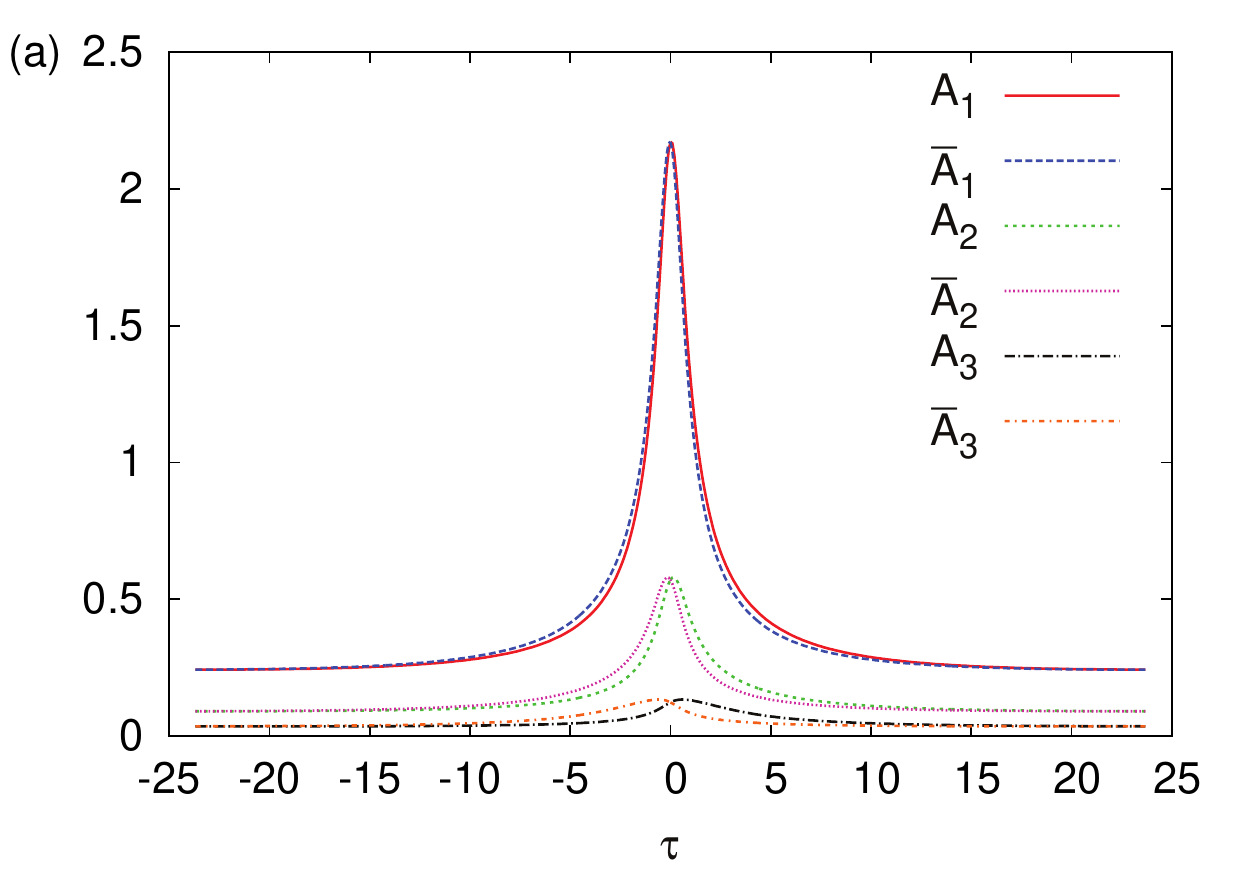}
\includegraphics[width=0.9\columnwidth]{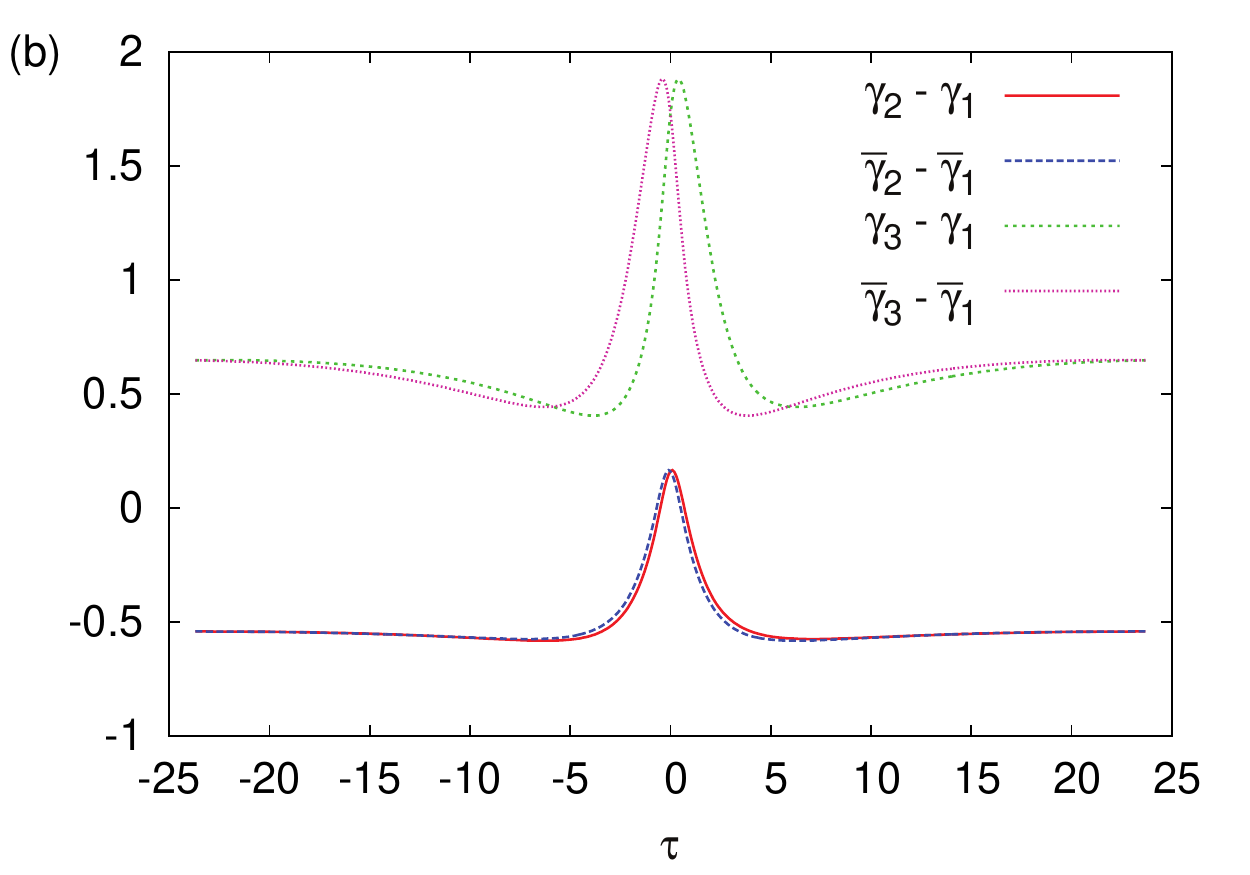}
 \caption{(Color online)
 Variational parameters (a) $A_k(\tau)$, $\bar A_k(\tau)$ for
 $k=1,\dots,N_g$ and (b) $\gamma_k(\tau)-\gamma_1(\tau)$,
 $\bar\gamma_k(\tau)-\bar\gamma_1(\tau)$ for $k=2,\dots,N_g$ of a
 periodic trajectory with period $\beta=47.4$ at scaled scattering
 length $a=-0.9$ described by $N_g=3$ coupled Gaussian functions.}
\label{fig2}
\end{figure}

As an example Fig.~\ref{fig2} shows the variational parameters of a
periodic trajectory at scaled scattering length $a=-0.9$ with period
$\beta=47.4$ near the bounce.
The BEC is described by a superposition of $N_g=3$ Gaussian
functions.
The high values of the parameters $A_k$ and $\bar A_k$ in
Fig.~\ref{fig2}(a) indicate a small width of the condensate wave
function at the bounce around $\tau=0$.

Once a bounce trajectory has been found improved values of the
parameters $S_{\rm b}$, $\omega_0$, and $v_0$ in the formula
\eqref{eq:Gamma0} for the decay rate of the BEC can be determined.
The Euclidean action $S_{\rm b}=\Delta S_{\ee}$ of a periodic trajectory
is given in Eq.~\eqref{eq:Se}.
For the ansatz with coupled Gaussians the action can directly be
expressed in terms of the variational parameters,
\begin{align}
 &\Delta S_{\ee} = \int_0^{\beta/2} \dd\tau \sum_{k,l=1}^{N_g} \nonumber \\ 
 &\times
   \left\{ \left(\dot{\bar A}_l - \dot A_k\right)
   \Braket{\bar g_l}{r^2}{g_k} + \left(\dot{\bar\gamma}_l-
   \dot\gamma_k\right) \langle\bar g_l|g_k\rangle \right\} \; .
\label{eq:Se_var}
\end{align}
The matrix elements are given in Eqs.~\eqref{eq:A1} and \eqref{eq:A2}
in Appendix~\ref{sec:app_A}.

The frequency $\omega_0$ is obtained, instead of using the potential
in Eq.~\eqref{eq:omega0}, by a stability analysis of the ground state.
For the ansatz with coupled Gaussian functions the stability
parameters are obtained as eigenvalues of the Jacobian matrix
constructed by varying the variational parameters around the fixed
point solution \cite{Rau10a}.
The eigenvalue $\omega_0$ belongs to the variation of the parameters
in the direction of the bounce trajectory.

The parameter $v_0$ cannot be determined using Eq.~\eqref{eq:v0_q}
because no canonical coordinate $q(\tau)$ exists for the ansatz with
coupled Gaussians.
However, when the bounce trajectory is approached via periodic orbits,
$v_0$ can be estimated using the mean-field energies $E_{\rm mf}^g$
and $E_{\rm mf}(\beta)$ of the ground state and periodic trajectories
with period $\beta$.
Using a model with an inverted harmonic potential for approaching the
saddle in imaginary time we obtain (see Appendix~\ref{sec:app_B})
\begin{equation}
 v_0 = \lim_{\beta\to\infty}
   \sqrt{\frac{E_{\rm mf}(\beta)-E_{\rm mf}^g}{2m}}\, \ee^{\omega_0\beta/2} \; .
\label{eq:v0}
\end{equation}
Results obtained with coupled Gaussians are presented in
Sec.~\ref{sec:res} and compared with numerically exact simulations on
space-time lattices.

\subsection{Numerical simulations on space-time lattices}
\label{sec:grid_calc}
To verify the validity of the extended variational approach with
coupled Gaussian functions we have carried out numerically exact
computations by simulating the bounce trajectory on a space-time
lattice.
Because the condensate wave function is spherically symmetric only the
$r$ coordinate needs to be discretized in space.
We set $\psi_{i,j}\equiv\psi(i\Delta r,j\Delta\tau)$ and
$\bar\psi_{i,j}\equiv\bar\psi(i\Delta r,j\Delta\tau)$ with $i=1,\dots,m$
and $j=1,\dots,n$ for an $(m,n)$ space-time lattice with step sizes
$\Delta r$ and $\Delta\tau$, respectively.

The propagation in imaginary time is obtained with the operators
\begin{equation}
 U = \ee^{-H\tau} \; , \quad \bar U = \ee^{H\tau} \; ,
\end{equation}
applied to the wave functions $\psi$ and $\bar\psi$, respectively.
The Hamiltonian $H=T+V_{\rm c}+V_{\rm u}-\mu$ is given in
Eq.~\eqref{eq:H}.
Because the operators of the kinetic energy $T=-\Delta$ and of the
potential $V=V_{\rm c}+V_{\rm u}$ do not commute we adopt the
Baker-Hausdorff formula
\begin{equation}
 \ee^{-(T+V)\Delta\tau} \approx
 \ee^{-V\Delta\tau/2}\, \ee^{-T\Delta\tau}\, \ee^{-V\Delta\tau/2} \; ,
\end{equation}
valid for small time steps $\Delta\tau$ and apply the split-operator
method.
The operators $\ee^{\pm V\Delta\tau/2}$ are applied in coordinate
space, and the operators $\ee^{\pm T\Delta\tau}$ in momentum space.
The Fourier transform of a spherically symmetric wave function
$\psi(r)$ yields
\begin{equation}
 \tilde\psi(k) = \mathcal{F}\{\psi(r)\}
  = \frac{4\pi}{k} \int_0^\infty r \psi(r) \sin(kr)\dd r \; ,
\end{equation}
which can be computed efficiently on the grid by a fast sine
transform \cite{NumRec}.
For the potential $V_{\rm u}(r)$ of the attractive $1/r$ interaction
[see Eq.~\eqref{eq:Vu_tau}] we obtain by application of the
convolution theorem
\begin{align}
 V_{\rm u}(r)
  &= \mathcal{F}^{-1}\left\{\mathcal{F}\{\bar\psi\psi\}
     \mathcal{F}\left\{\frac{1}{|\vec{r}|}\right\}\right\}
   = \frac{8}{r}\int_0^{\infty}\frac{1}{k^2} \nonumber \\
  &\times \left[\int_0^{\infty} r' \bar\psi(r')\psi(r')\sin(kr')
     \dd r'\right]\sin(kr)\dd k \; ,
\end{align}
where the integrals are also computed by fast sine transforms.
For large values of $r$ the potential $V_{\rm u}(r)$ asymptotically
approaches $-2/r$, however, the sine transform yields
$V_{\rm u}(r_{\max})=0$ at the border $r_{\max}=m\Delta r$ of the grid.
The error is corrected by a shift of the potential
$V_{\rm u}(r)\to V_{\rm u}(r)-2/r_{\max}$.

The space-time lattice can be initialized, e.g., by discretizing a
trajectory obtained with the extended variational approach introduced
in Sec.~\ref{sec:gauss_calc}.
The remaining task is then to fulfill the matching conditions that the
grid points of $\psi$ and $\bar\psi$ at time $\tau=j\Delta\tau$
propagated by a time step $\Delta\tau$ coincide with the grid points
of $\psi$ and $\bar\psi$ at time $\tau=(j+1)\Delta\tau$.
For a periodic trajectory furthermore the boundary conditions at the
turning points $\bar\psi(\tau=0)=\psi(\tau=0)$ and
$\bar\psi(\tau=n\Delta\tau)=\psi(\tau=n\Delta\tau)$ must be fulfilled.
The matching conditions yield a high-dimensional system of $2m(n-1)$
coupled equations, which can be solved by a Newton method.
For a typical size of the space-time lattice with $m=64$ and $n=151$
used in our calculations the Jacobian matrix in Newton's method has
the dimension $19200\times 19200$, which means about $3.7\times 10^8$
matrix elements.
However, the matching conditions only depend on the grid points at two
adjacent time steps, and thus the Jacobian has a well pronounced band
structure which allows for an efficient solution of the linear system
of equations \cite{NumRec}.
Furthermore, the norm conservation can be used to slightly reduce the
number of parameters.

Similar as discussed in Sec.~\ref{sec:gauss_calc} for the variational
calculations we start searching for periodic trajectories on the
space-time lattice with a short period $\beta$ at mean-field energies
slightly below the excited state (see Fig.~\ref{fig1}) and then
increase $\beta$.
\begin{figure}
\includegraphics[width=0.9\columnwidth]{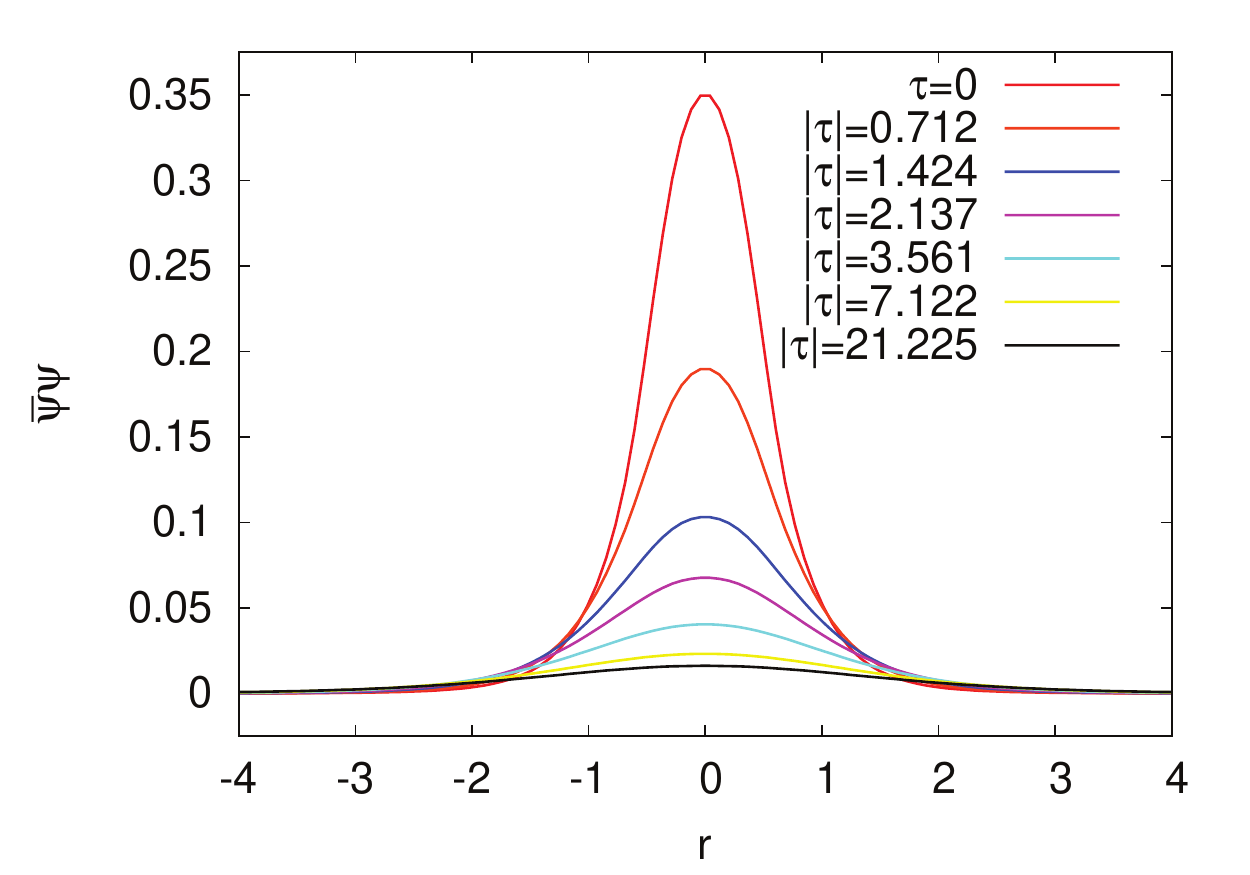}
\caption{(Color online)
 Density $[\bar\psi\psi](r)$ of a condensate wave function for a
 periodic trajectory with period $\beta=42.45$ computed on the
 space-time lattice at increasing values $|\tau|$ (from top to bottom
 line) of the imaginary time.
 The scaled scattering length is $a=-0.9$.}
\label{fig3}
\end{figure}
As an example Fig.~\ref{fig3} shows the density $[\bar\psi\psi](r)$ of
a condensate wave function for a periodic trajectory with period
$\beta=42.45$ computed on the space-time lattice at various values
of the imaginary time $\tau$.
The wave function is narrowest at the bounce at $\tau=0$ and becomes
broader with increasing time.
Note that the area under each curve in Fig.~\ref{fig3} is not the
square norm $2\pi\int_0^\infty r^2\bar\psi(r)\psi(r)\dd r$ and thus
not conserved.
As discussed above, the bounce trajectory is obtained in the limit
$\beta\to\infty$.

The Euclidean action of a periodic trajectory on the space-time
lattice is computed with Eq.~\eqref{eq:Se}.
The exact value of the frequency $\omega_0$ is obtained by numerically
solving the Bogoliubov-de Gennes equations at the stationary ground
state as discussed in Refs.~\cite{Car08b,Kre12}.
The parameter $v_0$ is determined with Eq.~\eqref{eq:v0}.

\section{Results and discussion}
\label{sec:res}
We now present and compare the results obtained with the extended
variational approach using coupled Gaussian functions and by the
numerically exact simulations on space-time lattices.
As outlined above the bounce trajectory is approached via periodic
trajectories fulfilling the boundary conditions $\bar\psi(0)=\psi(0)$
and $\bar\psi(\beta/2)=\psi(\beta/2)$ at the turning points in the
limit of an infinite period $\beta\to\infty$.
The periodic orbit search is started close to the stationary excited
state as illustrated in Fig.~\ref{fig1}(b) where the period
$\beta=2\pi/\omega_{\ee}$ is given by the corresponding stability
eigenvalue $\omega_{\ee}$ of that state.
The periodic orbit is then varied towards the bounce trajectory by
increasing $\beta$ in small steps.

\begin{figure}
\includegraphics[width=0.9\columnwidth]{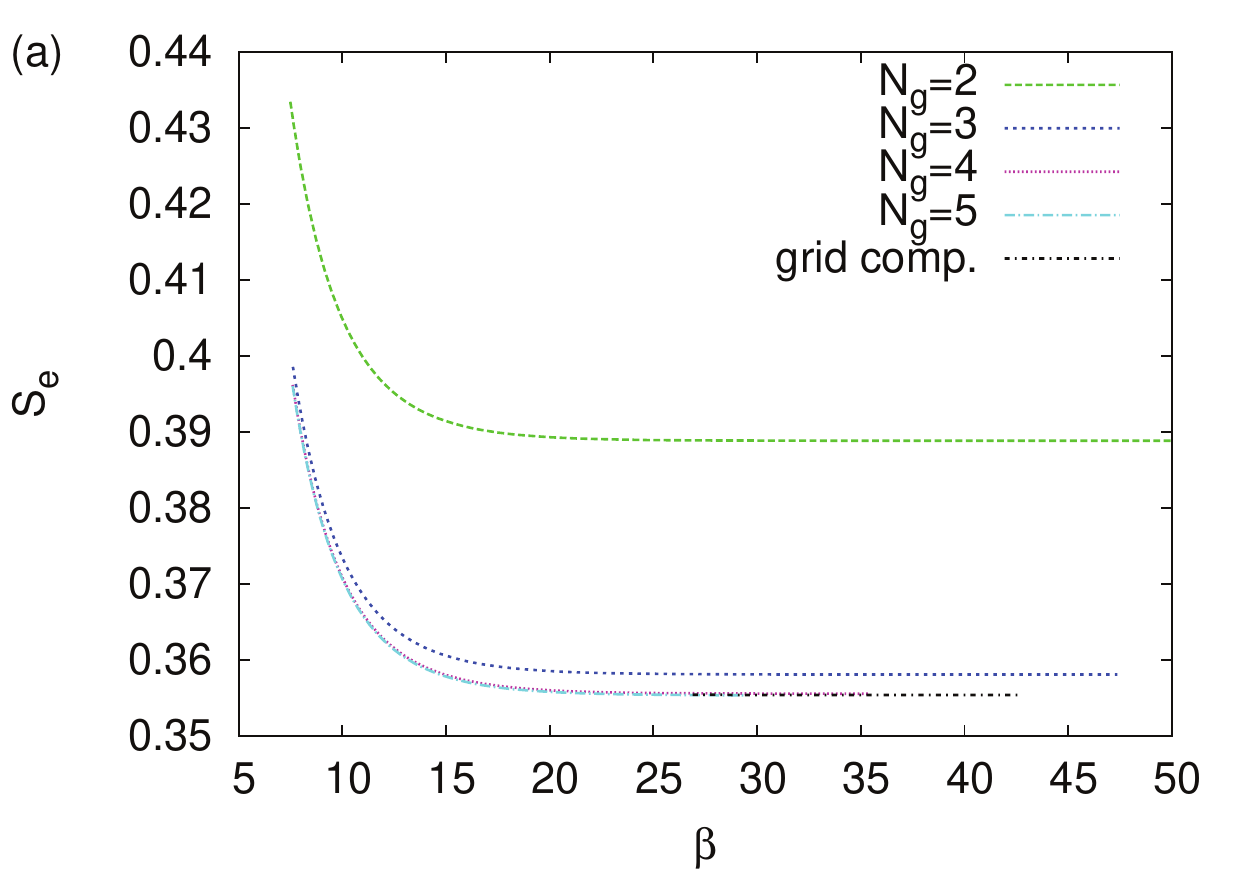}
\includegraphics[width=0.9\columnwidth]{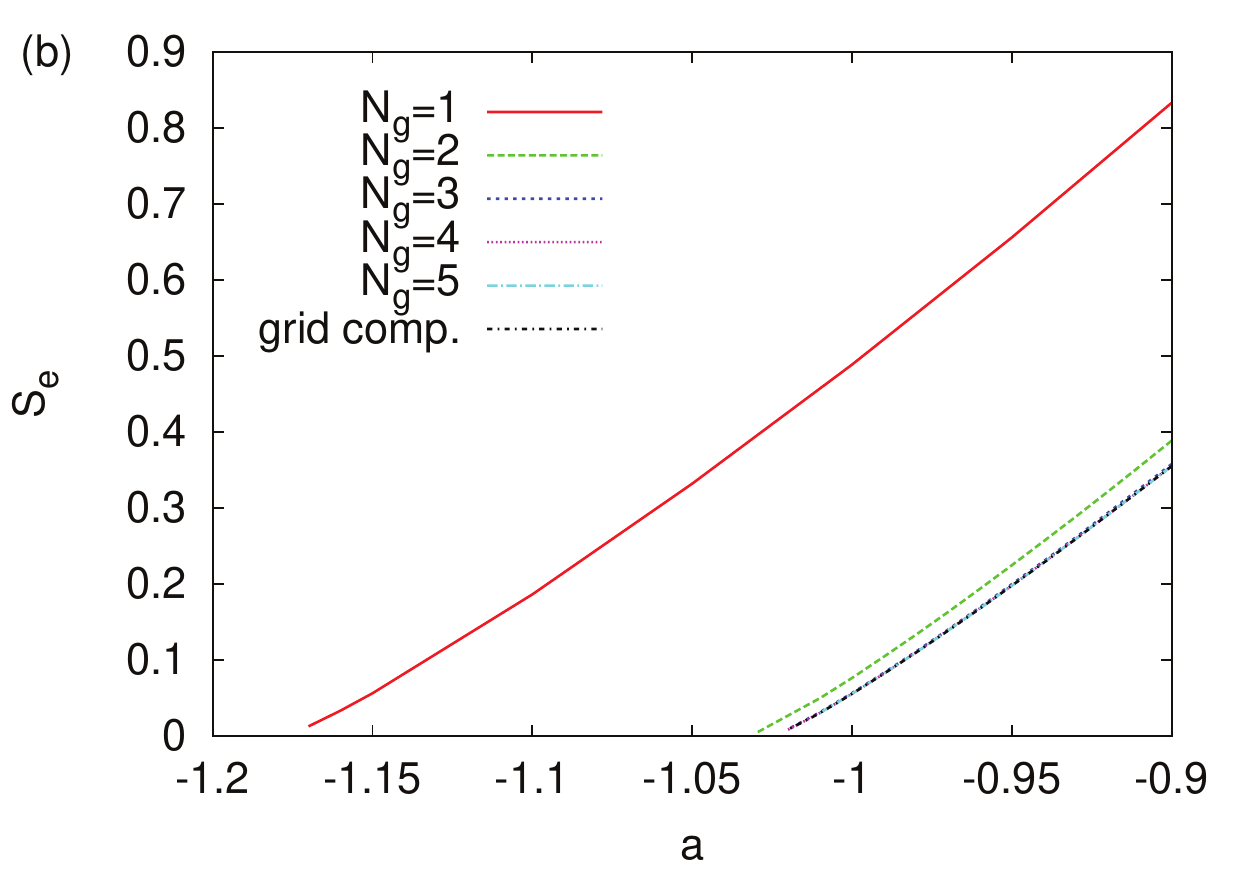}
\caption{(Color online)
 (a) Euclidean action of periodic trajectories at constant scaled
 scattering length $a=-0.9$.
 (b) Euclidean action of the bounce trajectory as function of the
 scaled scattering length $a$.  Shown are the results obtained by the
 variational approach using up to $N_g=5$ coupled Gaussians and by
 numerically exact simulations on grids.  The variational results with
 $N_g>3$ agree within the line width with the grid computations.}
\label{fig4}
\end{figure}

The Euclidean action of the periodic orbits at constant scaled
scattering length $a=-0.9$ as function of the period $\beta$ is shown
in Fig.~\ref{fig4}(a).
The results are obtained by the variational approach using up to
$N_g=5$ coupled Gaussians and by numerically exact simulations on
space-time lattices.
Evidently, the variational calculations converge rapidly to the
solution of the numerical grid calculations for an increasing number
of Gaussians.
For more than three Gaussians the deviations are less than the line
widths.
However, it is important to note that the converged action deviates
significantly from the result obtained with a single Gaussian function
[that action is $\sim 0.83$ for $a=-0.9$ and thus far outside the
region presented in Fig.~\ref{fig4}(a)].
The action decreases with increasing period, but becomes nearly
constant in the region $\beta\gtrsim 20$, which indicates the rapid
convergence to the action of the bounce trajectory.

The limit $\beta\to\infty$ of the Euclidean action has been estimated
for periodic trajectories at various scaled scattering lengths.
The resulting action of the bounce trajectory is presented in
Fig.~\ref{fig4}(b).
Again, the variational results converge rapidly to those obtained by
simulations on space-time lattices.
The action of the bounce trajectory vanishes at the critical
scattering length, where the ground and excited state emerge in a
tangent bifurcation.
The exact critical scattering length is $a_{\rm cr}^{\rm ex}=-1.0251$,
however, in the variational calculation with a single Gaussian the
bifurcation point is significantly shifted to
$a_{\rm cr}^{N_g=1}=-1.1781$.
As a consequence in the region $a>a_{\rm cr}^{\rm ex}=-1.0251$ the
difference between the action obtained with $N_g=1$ and the exact
action is $\sim 0.5$, i.e., quite large.

\begin{figure}
\includegraphics[width=0.9\columnwidth]{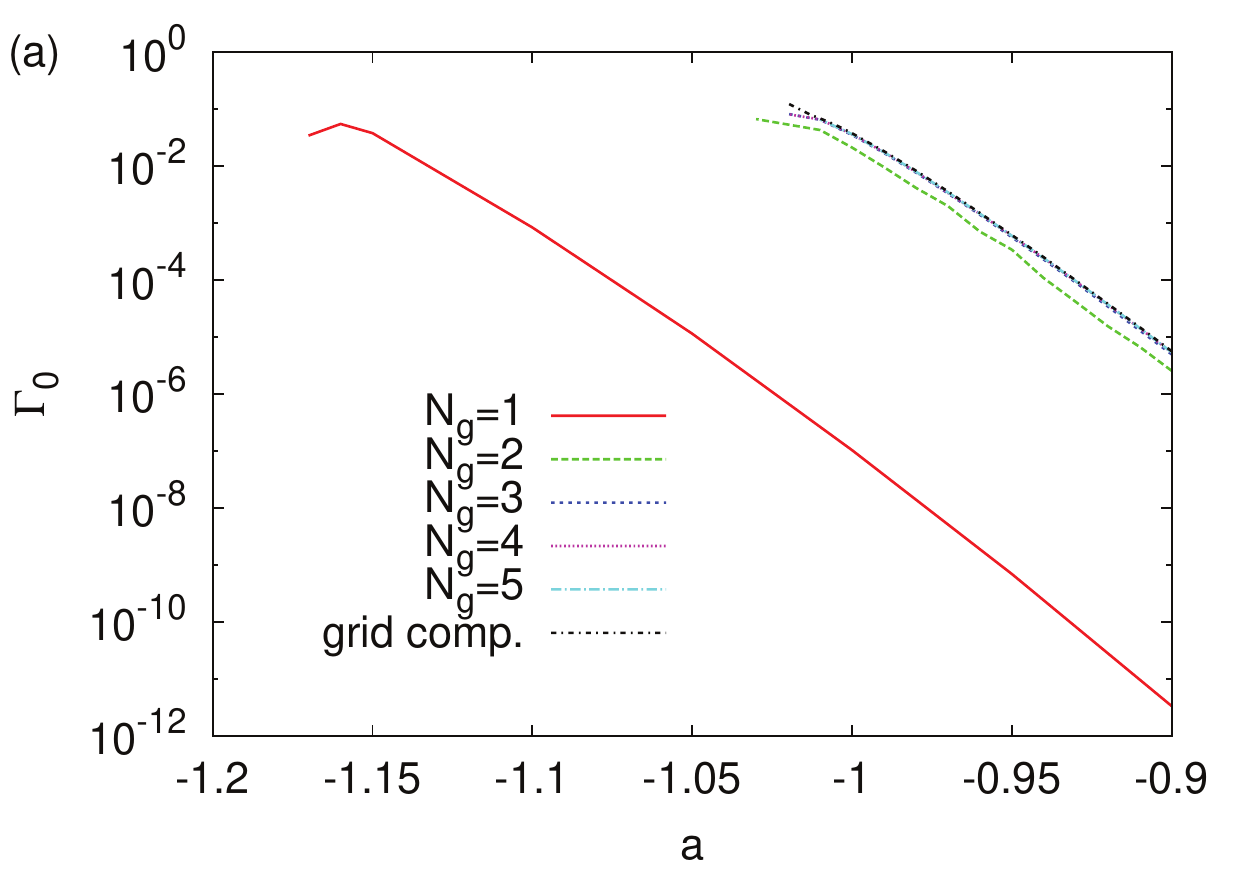}
\includegraphics[width=0.9\columnwidth]{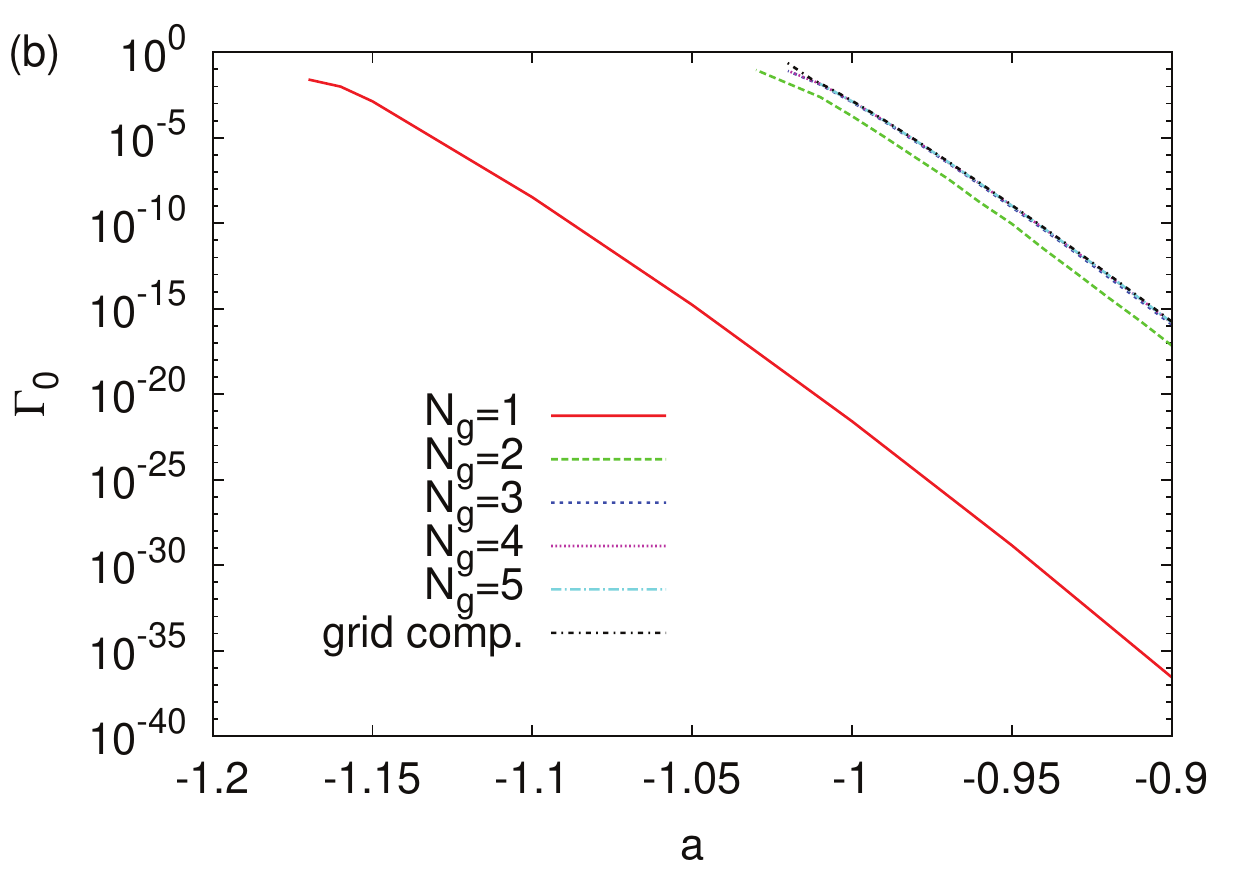}
\caption{(Color online)
 Decay rate by macroscopic quantum tunneling of a BEC with (a) $N=30$
 and (b) $N=100$ particles obtained by Eq.~\eqref{eq:Gamma0} with the
 parameters $S_{\rm b}$, $\omega_0$, and $v_0$ of the bounce trajectory.
 The variational calculations with $N_g=1$ to $5$ coupled Gaussians
 converge rapidly to the grid computations, for $N_g\ge 3$ deviations
 are less than the line width.}
\label{fig5}
\end{figure}
The Euclidean action shown in Fig.~\ref{fig4}(b) and the parameters
$\omega_0$ and $v_0$ of the bounce trajectory can now be inserted into
Eq.~\eqref{eq:Gamma0} to determine the decay rate $\Gamma_0$ of the
BEC with attractive $1/r$ interaction.
Although particle number scaled units [see Eq.~\eqref{eq:Nscal}] are
used in Eq.~\eqref{eq:Gamma0} the decay rate depends, through the
replacement $\hbar\to\hbar/N$ for the macroscopic quantum tunneling of
condensates \cite{Sto97}, explicitly on the particle number $N$.
As examples, results for a BEC with $N=30$ and $N=100$ bosons are
presented in Fig.~\ref{fig5}(a) and (b), respectively.
The decay rate in Eq.~\eqref{eq:Gamma0} is strongly dominated by the
exponential term $\exp(-NS_{\rm b}/\hbar)$.
The action of the bounce trajectory tends to zero at the critical
value $a_{\rm cr}$ of the tangent bifurcation [see
Fig.~\ref{fig4}(b)], and therefore macroscopic quantum tunneling can
play a non-negligible role at scattering lengths slightly above that
value.
The variational results in Fig.~\ref{fig5} converge rapidly to those
of the grid computations with increasing number of coupled Gaussians.
It should be noted that the decay rate obtained with a single Gaussian
function deviates in the region $a>a_{\rm cr}^{\rm ex}=-1.0251$ from
the converged result by about 12 orders of magnitude for $N=30$
and by about 20 orders of magnitude for $N=100$ particles.

The decay rates in Fig.~\ref{fig5} are given in particle number scaled
dimensionless units.
For the experiment proposed by O'Dell et al.\ \cite{ODe00}
experimentally feasible parameters have been found in Ref.~\cite{Gio01}.
With these values the scattering length and the decay rates can be
given in SI units.
For $^{87}$Rb with $a_u=\hbar^2/(mu)\sim 2.5\times 10^{-4}\,$m and
$t_u=2ma_u^2/\hbar\sim 27.1\,$s we present in Table~\ref{tab1} the decay
rates $\Gamma_0$ in s$^{-1}$ at scaled scattering length $a=-1$
corresponding to $a^{\rm (SI)}=-a_u/N^2=-2.78\times 10^{-7}\,$m for
$N=30$ and $a^{\rm (SI)}=-a_u/N^2=-2.5\times 10^{-8}\,$m for $N=100$
particles.
\begin{table}
\caption{Decay rates $\Gamma_0$ in SI units for self-trapped
  condensates of $N=30$ and $N=100$ $^{87}$Rb atoms with the
  parameters \cite{Gio01,Pap07}
  $a_u=\hbar^2/(mu)\sim 2.5\times 10^{-4}\,$m and
  $t_u=2ma_u^2/\hbar\sim 27.1\,$s.}
\begin{tabular}{c|l|l}
 $N_g$ & $\Gamma_0^{(N=30)}[{\rm s}^{-1}]$ & $\Gamma_0^{(N=100)}[{\rm s}^{-1}]$ \\
 \hline
 1 & $3.5\times 10^{-6}$  & $9.65\times 10^{-20}$ \\
 2 & $0.70$  & $0.066$ \\
 3 & $1.18$  & $0.45$ \\
 4 & $1.21$  & $0.49$ \\
 5 & $1.21$  & $0.50$ \\
 grid & $1.21$  & $0.50$
\end{tabular}
\label{tab1}
\end{table}
The values in Table~\ref{tab1} indicate that the lifetimes of the
self-trapped BEC with reasonable experimental parameters can be in the
order of seconds at scattering lengths close to the tangent
bifurcation.
The strong deviations between the decay rates obtained with a single
Gaussian function and with improved methods are also evident.
Thus, the results presented in Fig.~\ref{fig5} and Table~\ref{tab1}
clearly illustrate that the accurate computation of the bounce
trajectory is extremely important for a reasonable quantitative
description of macroscopic quantum tunneling.

\section{Conclusion and outlook}
\label{sec:conc}
We have shown that the exact computation of the bounce trajectory of a
BEC using an extended variational approach with coupled Gaussian
functions is a full-fledged alternative to numerically expensive
simulations on space-time lattices.
Very few (about three to five) Gaussians are sufficient to obtain
converged results.
The method here has been developed for a spherically symmetric
self-trapped BEC with an attractive $1/r$ interaction, however, it can
straightforwardly also be applied to condensates in an external trap
or with a different long-range interaction.
The variational approach may be of particular advantage for
interactions without spherical symmetry, e.g., in dipolar condensates,
which require an increased dimension of the space-time lattice for
numerical simulations.

The parameters of the exact bounce trajectory have been used as input
in the formula \eqref{eq:Gamma0} derived by Stoof \cite{Sto97}.
This certainly yields improved, however, still not exact values
$\Gamma_0$ for the decay rate of the BEC by macroscopic quantum
tunneling.
The reason is that the fluctuations around the stationary ground state
and the time-dependent bounce trajectory must be considered, i.e., the
decay rates $\Gamma_0$ in Eq.~\eqref{eq:Gamma0} must be multiplied with
the fluctuation determinant.
The computation of the fluctuation determinant and the investigation
of macroscopic quantum tunneling in dipolar condensates is work in
progress.

The discussions in this paper are based on the mean-field approach for
an effective one-particle condensate wave function.
For a BEC with attractive interactions and not too many particles
effects of depletion can, in principle, be studied beyond the
Gross-Pitaevskii theory.
In that case the dynamics of the decay should strongly depend on the
particle number, in particular for low numbers of particles, and the
loss of particles will probably increase the decay.
Such computations and comparisons with results obtained within the
mean-field approach are an interesting task for future work.

\acknowledgments
This work was supported by Deutsche Forschungsgemeinschaft.

\appendix

\section{Gaussian integrals}
\label{sec:app_A}
For the ansatz with real valued coupled Gaussian functions given in
Eq.~\eqref{eq:coupled_gauss} the integrals required in
Sec.~\ref{sec:gauss_calc} read
\begin{align}
\label{eq:A1}
 \langle\bar g_l|g_k\rangle
   &= \pi^{3/2} \frac{\ee^{-\gamma_{kl}}}{A_{kl}^{3/2}}
   \; ,\displaybreak[1]\\
\label{eq:A2}
 \Braket{\bar g_l}{r^2}{g_k}
   &= \frac{3}{2} \pi^{3/2} \frac{\ee^{-\gamma_{kl}}} {A_{kl}^{5/2}}
   \; ,\displaybreak[1]\\
 \Braket{\bar g_l}{r^4}{g_k}
   &= \frac{10}{3} \pi^{3/2} \frac{\ee^{-\gamma_{kl}} }{A_{kl}^{7/2}}
   \; ,\displaybreak[1]\\
 \Braket{\bar g_l}{{V}_{\mathrm{c}}}{g_k}
   &= 8\pi^{5/2} a \sum_{i,j=1}^{N_g} 
   \frac{\ee^{-\gamma_{ijkl}}}{\left(A_{ijkl}\right)^{3/2}}
   \; ,\displaybreak[1]\\
 \Braket{\bar g_l}{r^2 {V}_{\mathrm{c}}}{g_k}
   &= 12 \pi^{5/2} a \sum_{i,j=1}^{N_g}
   \frac{\ee^{-\gamma_{ijkl}}}{A_{ijkl}^{5/2}}
   \; ,\displaybreak[1]\\
 \Braket{\bar g_l}{{V}_{\mathrm{u}}}{g_k}
   &= - 4 \pi^{5/2} \sum_{i,j=1}^{N_g}
   \frac{\ee^{-\gamma_{ijkl}}}{ A_{ij} A_{kl} A_{ijkl}^{1/2}}
   \; ,\displaybreak[1]\\
 \Braket{\bar g_l}{r^2 {V}_{\mathrm{u}}}{g_k}
   &= -2\pi^{5/2} \sum_{i,j=1}^{N_g} \frac{\left(2 A_{ij} + 3 A_{kl}\right)
   \ee^{-\gamma_{ijkl}}}{ A_{ij}  A_{kl}^2 A_{ijkl}^{3/2}} \; ,
\end{align}
with
\begin{align}
 A_{kl} &= A_k + \bar A_l \; ,\\
 A_{ij} &= A_i + \bar A_j \; ,\displaybreak[0]\\
 A_{ijkl} &= A_i + \bar A_j + A_k + \bar A_l \; ,\displaybreak[0]\\
 \gamma_{kl} &= \gamma_k + \bar\gamma_l \; ,\\
 \gamma_{ijkl} &= \gamma_i + \bar\gamma_j + \gamma_k + \bar\gamma_l \; .
\end{align}

\section{Calculation of $v_0$}
\label{sec:app_B}
\begin{figure}
\includegraphics[width=0.8\columnwidth]{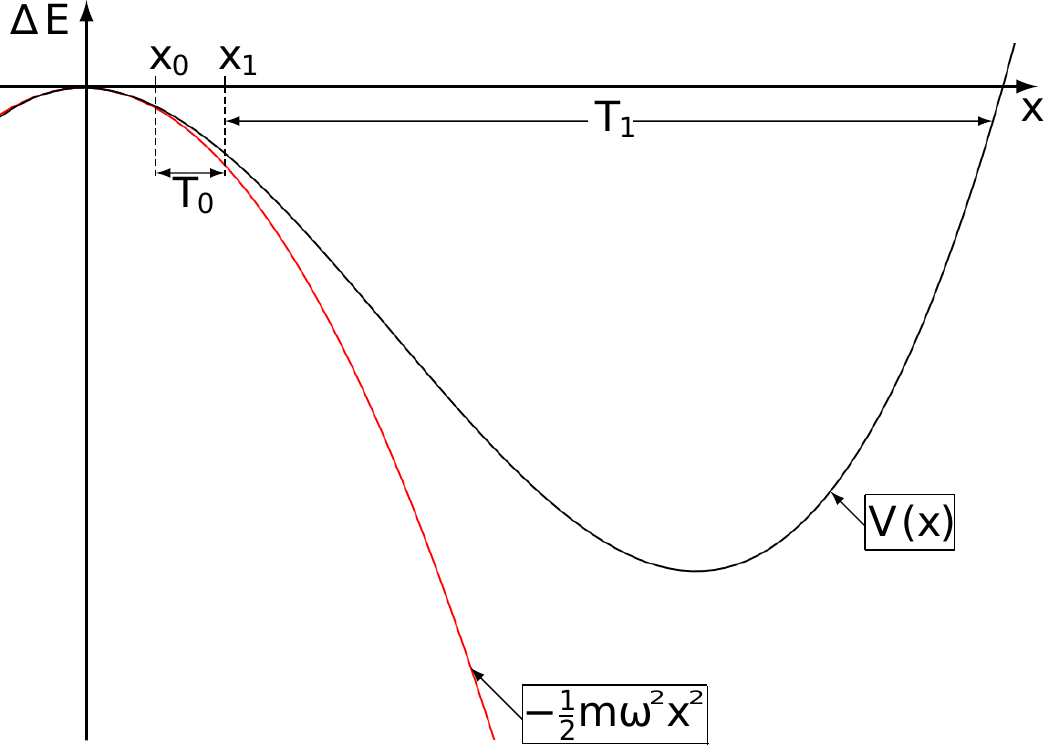}
\caption{(Color online)
 Sketch of a potential with an inverted harmonic saddle at the origin.}
\label{fig6}
\end{figure}

For the calculation of $v_0$ we assume a potential $V(x)$ with an
inverted harmonic saddle at the origin as illustrated in
Fig.~\ref{fig6}.
The period of a trajectory at energy $\Delta E<0$ is written as
$T(\Delta E)=2(T_0+T_1)$, with $T_0$ the time in the region $[x_0,x_1]$
where the harmonic approximation is valid (see Fig.~\ref{fig6}).
We obtain
\begin{align}
 T_0 &= \int_{x_0}^{x_1}\frac{\dd x}{\dot x} = \frac{1}{\omega}
   \int_{x_0}^{x_1} \frac{\dd x}{\sqrt{x^2 - x_0^2}} \nonumber \\
 &= \frac{1}{\omega} \cosh^{-1}{\frac{x_1}{x_0}} \overset{x_0 \ll x_1}{\approx}
 \frac{1}{\omega} \ln{\frac{2x_1}{x_0}} \nonumber \\
 \Rightarrow T({\Delta E}) &= 2(T_0+T_1) \approx \frac{2}{\omega}
 \ln{\left(\frac{2x_1}{x_0}\ee^{\omega T_1}\right)} \; .
\label{eq:B1}
\end{align}
When the turning point $x_0=\sqrt{-2\Delta E/m\omega^2}$ and the point
$x_1$ given by [see Eq.~\eqref{eq:v0_q}]
\begin{align}
 x_1 &= x(T_1) = \frac{v_0}{\omega} \ee^{-\omega T_1} \; ,
\label{eq:B2}
\end{align}
are inserted, Eq.~\eqref{eq:B1} yields
\begin{align}
  T({\Delta E})
  &\approx \frac{2}{\omega} \ln{\left(\frac{2 v_0}{\omega x_0}\right)} 
  = \frac{1}{\omega} \ln{\left( \frac{2 m v_0^2}{-{\Delta E}}\right)} 
 \nonumber \\
  \Rightarrow v_0 &= \sqrt{\frac{-\Delta E}{2m}} \ee^{\omega T/2} \; .
\label{eq:B3}
\end{align}
With $T=\beta$, $\omega=\omega_0$, $\Delta E=E_{\rm mf}^g-E_{\rm mf}(\beta)$,
and taking the limit $\beta\to\infty$ to approach the bounce
trajectory we end up with Eq.~\eqref{eq:v0}.


\begin{thebibliography}{10}%
\makeatletter
\providecommand \@ifxundefined [1]{%
 \ifx #1\undefined \expandafter \@firstoftwo
 \else \expandafter \@secondoftwo
\fi
}%
\providecommand \@ifnum [1]{%
 \ifnum #1\expandafter \@firstoftwo
 \else \expandafter \@secondoftwo
\fi
}%
\providecommand \enquote [1]{``#1''}%
\providecommand \bibnamefont  [1]{#1}%
\providecommand \bibfnamefont [1]{#1}%
\providecommand \citenamefont [1]{#1}%
\providecommand\href[0]{\@sanitize\@href}%
\providecommand\@href[1]{\endgroup\@@startlink{#1}\endgroup\@@href}%
\providecommand\@@href[1]{#1\@@endlink}%
\providecommand \@sanitize [0]{\begingroup\catcode`\&12\catcode`\#12\relax}%
\@ifxundefined \pdfoutput {\@firstoftwo}{%
 \@ifnum{\z@=\pdfoutput}{\@firstoftwo}{\@secondoftwo}%
}{%
 \providecommand\@@startlink[1]{\leavevmode}%
 \providecommand\@@endlink[0]{}%
}{%
 \providecommand\@@startlink[1]{%
  \leavevmode
  \pdfstartlink
   attr{/Border[0 0 1 ]/H/I/C[0 1 1]}%
   user{/Subtype/Link/A<</Type/Action/S/URI/URI(#1)>>}%
  \relax
 }%
 \providecommand\@@endlink[0]{\pdfendlink}%
}%
\providecommand \url  [0]{\begingroup\@sanitize \@url }%
\providecommand \@url [1]{\endgroup\@href {#1}{\urlprefix}}%
\providecommand \urlprefix [0]{URL }%
\providecommand \Eprint[0]{\href }%
\@ifxundefined \urlstyle {%
  \providecommand \doi [1]{doi:\discretionary{}{}{}#1}%
}{%
  \providecommand \doi [0]{doi:\discretionary{}{}{}\begingroup
  \urlstyle{rm}\Url }%
}%
\providecommand \doibase [0]{http://dx.doi.org/}%
\providecommand \Doi[1]{\href{\doibase#1}}%
\providecommand \bibAnnote [3]{%
  \BibitemShut{#1}%
  \begin{quotation}\noindent
    \textsc{Key:}\ #2\\\textsc{Annotation:}\ #3%
  \end{quotation}%
}%
\providecommand \bibAnnoteFile [2]{%
  \IfFileExists{#2}{\bibAnnote {#1} {#2} {\input{#2}}}{}%
}%
\providecommand \typeout [0]{\immediate \write \m@ne }%
\providecommand \selectlanguage [0]{\@gobble}%
\providecommand \bibinfo [0]{\@secondoftwo}%
\providecommand \bibfield [0]{\@secondoftwo}%
\providecommand \translation [1]{[#1]}%
\providecommand \BibitemOpen[0]{}%
\providecommand \bibitemStop [0]{}%
\providecommand \bibitemNoStop [0]{.\EOS\space}%
\providecommand \EOS [0]{\spacefactor3000\relax}%
\providecommand \BibitemShut [1]{\csname bibitem#1\endcsname}%
\bibitem{And95a}%
  \BibitemOpen
  \bibfield{author}{%
  \bibinfo {author} {\bibfnamefont{M.~H.}\ \bibnamefont{Anderson}}, \bibinfo
  {author} {\bibfnamefont{J.~R.}\ \bibnamefont{Ensher}}, \bibinfo {author}
  {\bibfnamefont{M.~R.}\ \bibnamefont{Matthews}}, \bibinfo {author}
  {\bibfnamefont{C.~E.}\ \bibnamefont{Wieman}},\ and\ \bibinfo {author}
  {\bibfnamefont{E.~A.}\ \bibnamefont{Cornell}},\ }%
  \bibfield{journal}{%
  \bibinfo {journal} {Science}\ }%
  \textbf{\bibinfo {volume} {269}},\ \bibinfo {pages} {198} (\bibinfo {year}
  {1995})%
  \bibAnnoteFile{NoStop}{And95a}%
\bibitem{Bra95a}%
  \BibitemOpen
  \bibfield{author}{%
  \bibinfo {author} {\bibfnamefont{C.~C.}\ \bibnamefont{Bradley}}, \bibinfo
  {author} {\bibfnamefont{C.~A.}\ \bibnamefont{Sackett}}, \bibinfo {author}
  {\bibfnamefont{J.~J.}\ \bibnamefont{Tollett}},\ and\ \bibinfo {author}
  {\bibfnamefont{R.~G.}\ \bibnamefont{Hulet}},\ }%
  \bibfield{journal}{%
  \Doi{10.1103/PhysRevLett.75.1687}{\bibinfo {journal} {Phys. Rev. Lett.}}\ }%
  \textbf{\bibinfo {volume} {75}},\ \bibinfo {pages} {1687} (\bibinfo {year}
  {1995})%
  \bibAnnoteFile{NoStop}{Bra95a}%
\bibitem{Dav95a}%
  \BibitemOpen
  \bibfield{author}{%
  \bibinfo {author} {\bibfnamefont{K.~B.}\ \bibnamefont{Davis}}, \bibinfo
  {author} {\bibfnamefont{M.~O.}\ \bibnamefont{Mewes}}, \bibinfo {author}
  {\bibfnamefont{M.~R.}\ \bibnamefont{Andrews}}, \bibinfo {author}
  {\bibfnamefont{N.~J.}\ \bibnamefont{van Druten}}, \bibinfo {author}
  {\bibfnamefont{D.~S.}\ \bibnamefont{Durfee}}, \bibinfo {author}
  {\bibfnamefont{D.~M.}\ \bibnamefont{Kurn}},\ and\ \bibinfo {author}
  {\bibfnamefont{W.}~\bibnamefont{Ketterle}},\ }%
  \bibfield{journal}{%
  \Doi{10.1103/PhysRevLett.75.3969}{\bibinfo {journal} {Phys. Rev. Lett.}}\ }%
  \textbf{\bibinfo {volume} {75}},\ \bibinfo {pages} {3969} (\bibinfo {year}
  {1995})%
  \bibAnnoteFile{NoStop}{Dav95a}%
\bibitem{Lahaye09}%
  \BibitemOpen
  \bibfield{author}{%
  \bibinfo {author} {\bibfnamefont{T.}~\bibnamefont{Lahaye}}, \bibinfo {author}
  {\bibfnamefont{C.}~\bibnamefont{Menotti}}, \bibinfo {author}
  {\bibfnamefont{L.}~\bibnamefont{Santos}}, \bibinfo {author}
  {\bibfnamefont{M.}~\bibnamefont{Lewenstein}},\ and\ \bibinfo {author}
  {\bibfnamefont{T.}~\bibnamefont{Pfau}},\ }%
  \bibfield{journal}{%
  \bibinfo {journal} {Rep. Prog. Phys.}\ }%
  \textbf{\bibinfo {volume} {72}},\ \bibinfo {pages} {126401} (\bibinfo {year}
  {2009})%
  \bibAnnoteFile{NoStop}{Lahaye09}%
\bibitem{ODe00}%
  \BibitemOpen
  \bibfield{author}{%
  \bibinfo {author} {\bibfnamefont{D.}~\bibnamefont{O'Dell}}, \bibinfo {author}
  {\bibfnamefont{S.}~\bibnamefont{Giovanazzi}}, \bibinfo {author}
  {\bibfnamefont{G.}~\bibnamefont{Kurizki}},\ and\ \bibinfo {author}
  {\bibfnamefont{V.~M.}\ \bibnamefont{Akulin}},\ }%
  \bibfield{journal}{%
  \Doi{10.1103/physrevlett.84.5687}{\bibinfo {journal} {Phys. Rev. Lett.}}\ }%
  \textbf{\bibinfo {volume} {84}},\ \bibinfo {pages} {5687} (\bibinfo {year}
  {2000})%
  \bibAnnoteFile{NoStop}{ODe00}%
\bibitem{Pap07}%
  \BibitemOpen
  \bibfield{author}{%
  \bibinfo {author} {\bibfnamefont{I.}~\bibnamefont{Papadopoulos}}, \bibinfo
  {author} {\bibfnamefont{P.}~\bibnamefont{Wagner}}, \bibinfo {author}
  {\bibfnamefont{G.}~\bibnamefont{Wunner}},\ and\ \bibinfo {author}
  {\bibfnamefont{J.}~\bibnamefont{Main}},\ }%
  \bibfield{journal}{%
  \Doi{10.1103/PhysRevA.76.053604}{\bibinfo {journal} {Phys. Rev. A}}\ }%
  \textbf{\bibinfo {volume} {76}},\ \bibinfo {eid} {053604} (\bibinfo {year}
  {2007})%
  \bibAnnoteFile{NoStop}{Pap07}%
\bibitem{Car08a}%
  \BibitemOpen
  \bibfield{author}{%
  \bibinfo {author} {\bibfnamefont{H.}~\bibnamefont{Cartarius}}, \bibinfo
  {author} {\bibfnamefont{J.}~\bibnamefont{Main}},\ and\ \bibinfo {author}
  {\bibfnamefont{G.}~\bibnamefont{Wunner}},\ }%
  \bibfield{journal}{%
  \Doi{10.1103/PhysRevA.77.013618}{\bibinfo {journal} {Phys. Rev. A}}\ }%
  \textbf{\bibinfo {volume} {77}},\ \bibinfo {pages} {013618} (\bibinfo {year}
  {2008})%
  \bibAnnoteFile{NoStop}{Car08a}%
\bibitem{Car08b}%
  \BibitemOpen
  \bibfield{author}{%
  \bibinfo {author} {\bibfnamefont{H.}~\bibnamefont{Cartarius}}, \bibinfo
  {author} {\bibfnamefont{T.}~\bibnamefont{Fab\v{c}i\v{c}}}, \bibinfo {author}
  {\bibfnamefont{J.}~\bibnamefont{Main}},\ and\ \bibinfo {author}
  {\bibfnamefont{G.}~\bibnamefont{Wunner}},\ }%
  \bibfield{journal}{%
  \Doi{10.1103/PhysRevA.78.013615}{\bibinfo {journal} {Phys. Rev. A}}\ }%
  \textbf{\bibinfo {volume} {78}},\ \bibinfo {pages} {013615} (\bibinfo {year}
  {2008})%
  \bibAnnoteFile{NoStop}{Car08b}%
\bibitem{Hue99}%
  \BibitemOpen
  \bibfield{author}{%
  \bibinfo {author} {\bibfnamefont{C.}~\bibnamefont{Huepe}}, \bibinfo {author}
  {\bibfnamefont{S.}~\bibnamefont{M{\'e}tens}}, \bibinfo {author}
  {\bibfnamefont{G.}~\bibnamefont{Dewel}}, \bibinfo {author}
  {\bibfnamefont{P.}~\bibnamefont{Borckmans}},\ and\ \bibinfo {author}
  {\bibfnamefont{M.~E.}\ \bibnamefont{Brachet}},\ }%
  \bibfield{journal}{%
  \Doi{10.1103/PhysRevLett.82.1616}{\bibinfo {journal} {Phys. Rev. Lett.}}\ }%
  \textbf{\bibinfo {volume} {82}},\ \bibinfo {pages} {1616} (\bibinfo {year}
  {1999})%
  \bibAnnoteFile{NoStop}{Hue99}%
\bibitem{Hue03}%
  \BibitemOpen
  \bibfield{author}{%
  \bibinfo {author} {\bibfnamefont{C.}~\bibnamefont{Huepe}}, \bibinfo {author}
  {\bibfnamefont{L.~S.}\ \bibnamefont{Tuckerman}}, \bibinfo {author}
  {\bibfnamefont{S.}~\bibnamefont{M{\'e}tens}},\ and\ \bibinfo {author}
  {\bibfnamefont{M.~E.}\ \bibnamefont{Brachet}},\ }%
  \bibfield{journal}{%
  \Doi{10.1103/PhysRevA.68.023609}{\bibinfo {journal} {Phys. Rev. A}}\ }%
  \textbf{\bibinfo {volume} {68}},\ \bibinfo {pages} {023609} (\bibinfo {year}
  {2003})%
  \bibAnnoteFile{NoStop}{Hue03}%
\bibitem{Jun12b}%
  \BibitemOpen
  \bibfield{author}{%
  \bibinfo {author} {\bibfnamefont{A.}~\bibnamefont{Junginger}}, \bibinfo
  {author} {\bibfnamefont{M.}~\bibnamefont{Dorwarth}}, \bibinfo {author}
  {\bibfnamefont{J.}~\bibnamefont{Main}},\ and\ \bibinfo {author}
  {\bibfnamefont{G.}~\bibnamefont{Wunner}},\ }%
  \bibfield{journal}{%
  \bibinfo {journal} {J. Phys. A: Math. Theor.}\ }%
  \textbf{\bibinfo {volume} {45}},\ \bibinfo {pages} {155202} (\bibinfo {year}
  {2012})%
  \bibAnnoteFile{NoStop}{Jun12b}%
\bibitem{Jun12c}%
  \BibitemOpen
  \bibfield{author}{%
  \bibinfo {author} {\bibfnamefont{A.}~\bibnamefont{Junginger}}, \bibinfo
  {author} {\bibfnamefont{J.}~\bibnamefont{Main}}, \bibinfo {author}
  {\bibfnamefont{G.}~\bibnamefont{Wunner}},\ and\ \bibinfo {author}
  {\bibfnamefont{T.}~\bibnamefont{Bartsch}},\ }%
  \bibfield{journal}{%
  \Doi{10.1103/PhysRevA.86.023632}{\bibinfo {journal} {Phys. Rev. A}}\ }%
  \textbf{\bibinfo {volume} {86}},\ \bibinfo {pages} {023632} (\bibinfo {year}
  {2012})%
  \bibAnnoteFile{NoStop}{Jun12c}%
\bibitem{Sto97}%
  \BibitemOpen
  \bibfield{author}{%
  \bibinfo {author} {\bibfnamefont{H.}~\bibnamefont{Stoof}},\ }%
  \bibfield{journal}{%
  \Doi{10.1007/bf02181289}{\bibinfo {journal} {J. Stat. Phys.}}\ }%
  \textbf{\bibinfo {volume} {87}},\ \bibinfo {pages} {1353} (\bibinfo {year}
  {1997})%
  \bibAnnoteFile{NoStop}{Sto97}%
\bibitem{Fre99}%
  \BibitemOpen
  \bibfield{author}{%
  \bibinfo {author} {\bibfnamefont{J.~A.}\ \bibnamefont{Freire}}\ and\ \bibinfo
  {author} {\bibfnamefont{D.~P.}\ \bibnamefont{Arovas}},\ }%
  \bibfield{journal}{%
  \Doi{10.1103/PhysRevA.59.1461}{\bibinfo {journal} {Phys. Rev. A}}\ }%
  \textbf{\bibinfo {volume} {59}},\ \bibinfo {pages} {1461 } (\bibinfo {year}
  {1999})%
  \bibAnnoteFile{NoStop}{Fre99}%
\bibitem{Rau10a}%
  \BibitemOpen
  \bibfield{author}{%
  \bibinfo {author} {\bibfnamefont{S.}~\bibnamefont{Rau}}, \bibinfo {author}
  {\bibfnamefont{J.}~\bibnamefont{Main}},\ and\ \bibinfo {author}
  {\bibfnamefont{G.}~\bibnamefont{Wunner}},\ }%
  \bibfield{journal}{%
  \Doi{10.1103/PhysRevA.82.023610}{\bibinfo {journal} {Phys. Rev. A}}\ }%
  \textbf{\bibinfo {volume} {82}},\ \bibinfo {pages} {023610} (\bibinfo {year}
  {2010})%
  \bibAnnoteFile{NoStop}{Rau10a}%
\bibitem{Rau10b}%
  \BibitemOpen
  \bibfield{author}{%
  \bibinfo {author} {\bibfnamefont{S.}~\bibnamefont{Rau}}, \bibinfo {author}
  {\bibfnamefont{J.}~\bibnamefont{Main}}, \bibinfo {author}
  {\bibfnamefont{H.}~\bibnamefont{Cartarius}}, \bibinfo {author}
  {\bibfnamefont{P.}~\bibnamefont{K{\"o}berle}},\ and\ \bibinfo {author}
  {\bibfnamefont{G.}~\bibnamefont{Wunner}},\ }%
  \bibfield{journal}{%
  \Doi{10.1103/PhysRevA.82.023611}{\bibinfo {journal} {Phys. Rev. A}}\ }%
  \textbf{\bibinfo {volume} {82}},\ \bibinfo {pages} {023611} (\bibinfo {year}
  {2010})%
  \bibAnnoteFile{NoStop}{Rau10b}%
\bibitem{Fre97}%
  \BibitemOpen
  \bibfield{author}{%
  \bibinfo {author} {\bibfnamefont{J.~A.}\ \bibnamefont{Freire}}, \bibinfo
  {author} {\bibfnamefont{D.~P.}\ \bibnamefont{Arovas}},\ and\ \bibinfo
  {author} {\bibfnamefont{H.}~\bibnamefont{Levine}},\ }%
  \bibfield{journal}{%
  \Doi{10.1103/PhysRevLett.79.5054}{\bibinfo {journal} {Phys. Rev. Lett.}}\ }%
  \textbf{\bibinfo {volume} {79}},\ \bibinfo {pages} {5054} (\bibinfo {year}
  {1997})%
  \bibAnnoteFile{NoStop}{Fre97}%
\bibitem{Pit03}%
  \BibitemOpen
  \bibfield{author}{%
  \bibinfo {author} {\bibfnamefont{L.~P.}\ \bibnamefont{Pitaevskii}}\ and\
  \bibinfo {author} {\bibfnamefont{S.}~\bibnamefont{Stringari}},\ }%
  \emph{\bibinfo {title} {{Bose-Einstein Condensation}}}\ (\bibinfo {publisher}
  {Oxford University Press},\ \bibinfo {year} {2003})%
  \bibAnnoteFile{NoStop}{Pit03}%
\bibitem{Dir30}%
  \BibitemOpen
  \bibfield{author}{%
  \bibinfo {author} {\bibfnamefont{P.~A.~M.}\ \bibnamefont{Dirac}},\ }%
  \bibfield{journal}{%
  \bibinfo {journal} {Math. Proc. Cam. Phil. Soc.}\ }%
  \textbf{\bibinfo {volume} {26}},\ \bibinfo {pages} {376} (\bibinfo {year}
  {1930})%
  \bibAnnoteFile{NoStop}{Dir30}%
\bibitem{McL64}%
  \BibitemOpen
  \bibfield{author}{%
  \bibinfo {author} {\bibfnamefont{A.~D.}\ \bibnamefont{McLachlan}},\ }%
  \bibfield{journal}{%
  \bibinfo {journal} {Mol. Phys.}\ }%
  \textbf{\bibinfo {volume} {8}},\ \bibinfo {pages} {39} (\bibinfo {year}
  {1964})%
  \bibAnnoteFile{NoStop}{McL64}%
\bibitem{NumRec}%
  \BibitemOpen
  \bibfield{author}{%
  \bibinfo {author} {\bibfnamefont{W.~H.}\ \bibnamefont{Press}}, \bibinfo
  {author} {\bibfnamefont{S.~A.}\ \bibnamefont{Teukolsky}}, \bibinfo {author}
  {\bibfnamefont{W.~T.}\ \bibnamefont{Vetterling}},\ and\ \bibinfo {author}
  {\bibfnamefont{B.~P.}\ \bibnamefont{Flannery}},\ }%
  \emph{\bibinfo {title} {Numerical Recipes in Fortran, Second Edition}}\
  (\bibinfo {publisher} {Cambridge University Press},\ \bibinfo {address}
  {Cambridge},\ \bibinfo {year} {1992})%
  \bibAnnoteFile{NoStop}{NumRec}%
\bibitem{Kre12}%
  \BibitemOpen
  \bibfield{author}{%
  \bibinfo {author} {\bibfnamefont{M.}~\bibnamefont{Kreibich}}, \bibinfo
  {author} {\bibfnamefont{J.}~\bibnamefont{Main}},\ and\ \bibinfo {author}
  {\bibfnamefont{G.}~\bibnamefont{Wunner}},\ }%
  \bibfield{journal}{%
  \Doi{10.1103/PhysRevA.86.013608}{\bibinfo {journal} {Phys. Rev. A}}\ }%
  \textbf{\bibinfo {volume} {86}},\ \bibinfo {pages} {013608} (\bibinfo {year}
  {2012})%
  \bibAnnoteFile{NoStop}{Kre12}%
\bibitem{Gio01}%
  \BibitemOpen
  \bibfield{author}{%
  \bibinfo {author} {\bibfnamefont{S.}~\bibnamefont{Giovanazzi}}, \bibinfo
  {author} {\bibfnamefont{D.}~\bibnamefont{O'Dell}},\ and\ \bibinfo {author}
  {\bibfnamefont{G.}~\bibnamefont{Kurizki}},\ }%
  \bibfield{journal}{%
  \Doi{10.1103/PhysRevA.63.031603}{\bibinfo {journal} {Phys. Rev. A}}\ }%
  \textbf{\bibinfo {volume} {63}},\ \bibinfo {pages} {031603(R)} (\bibinfo
  {year} {2001})%
  \bibAnnoteFile{NoStop}{Gio01}%
\end{thebibliography}
%

\end{document}